\documentstyle[11pt,newpasp,twoside,epsf]{article}
\def\edcomment#1{\iffalse\marginpar{\raggedright\sl#1\/}\else\relax\fi}
\reversemarginpar

\begin{document}

\title{The evolution of the Milky Way monitored in the solar neighbourhood}
\author{B. Fuchs, C. Dettbarn, H. Jahrei{\ss}, R. Wielen}

\affil{
Astronomisches Rechen-Institut,
M\"onchhofstrasse 12 - 14,\\ 69120 Heidelberg, Germany}


\begin{abstract}
In this review we concentrate on the dynamical evolution of the Milky Way as
monitored in the solar neighbourhood. The relevant data sets are presented
and discussed in detail. In the second part we review various mechanisms,
which drive the dynamical evolution.
\end{abstract}

\section{
Dynamical evolution of the galactic disk traced in the solar\\ neighbourhood}

The key data set, which we use for this study, is the Fourth Catalogue of
Nearby Stars (hereafter referred to as CNS4), which has been now completed
in its preliminary form (cf.~Jahrei{\ss} et al.~1998, 1999). The catalogue
represents the most complete inventory of stars within a distance of 25 pc
from the Sun. It contains data of about 3000 stars, most of them improved by
Hipparcos (ESA 1997) parallaxes. For instance, the parallaxes of 1411 stars
out of the 1761 stars with parallaxes more accurate than 10\% are Hipparcos
measurements. Statistical tests have shown that the CNS4 is complete for stars
with absolute magnitudes $M_{\rm V} \leq 8^m$, which includes the K stars. For
M stars the CNS4 becomes increasingly incomplete and kinematically
biased towards high proper motions.
\begin{figure} [h]
\begin{center}
\epsfxsize=9cm
   \leavevmode
     \epsffile{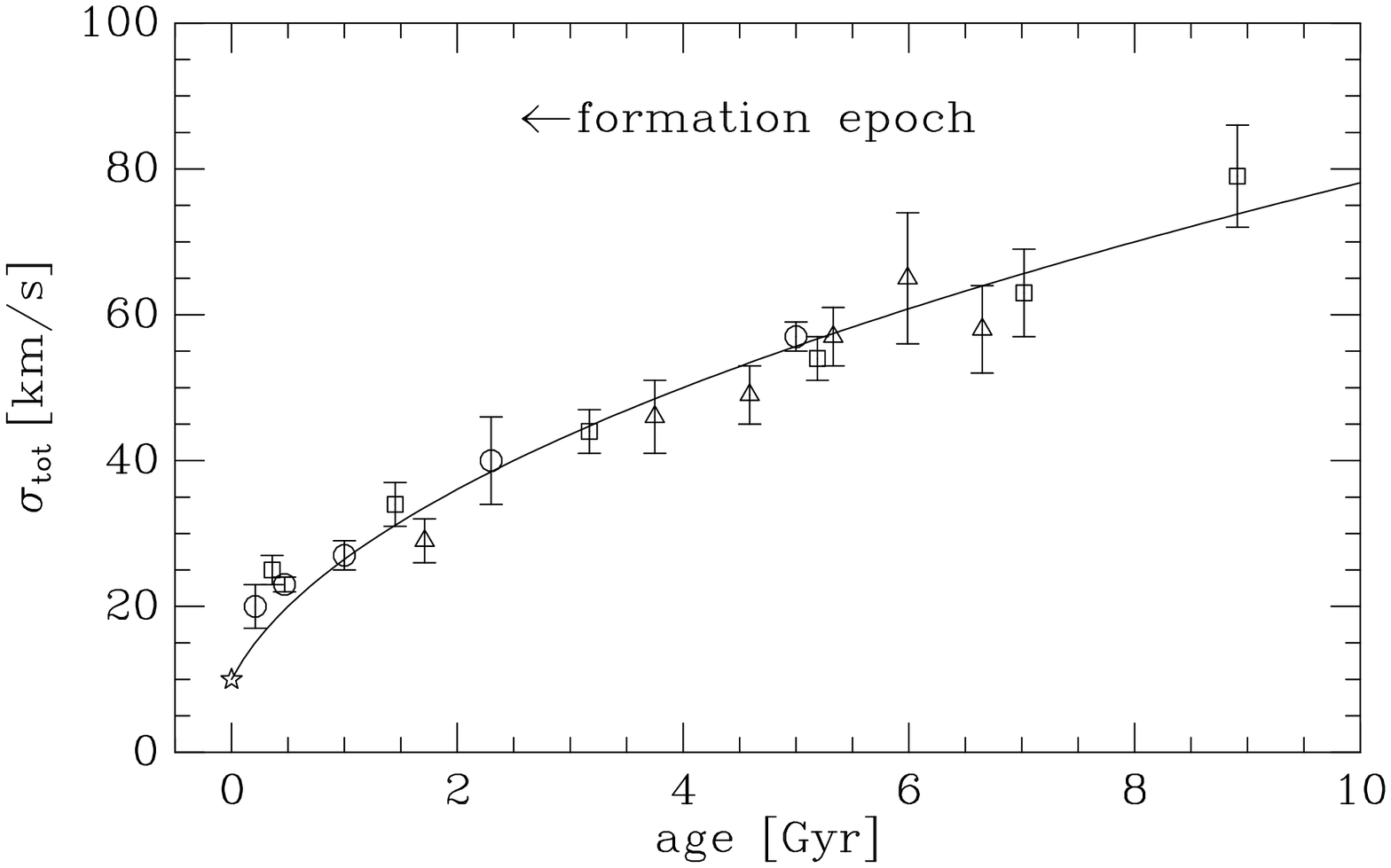}
\caption{Total velocity dispersions of the peculiar velocities of stars in the
solar neighbourhood. Circles indicate main sequence stars, squares McCormick
stars, and triangles refer to G and K dwarfs with quantitatively measured
chromospheric emission fluxes. Cepheids are indicated by an asterisk.}
\label{fig1}
   \end{center}
   \end{figure}

Fig.~1 illustrates the kinematical evolution of the galactic disk, where the
total velocity dispersions of the peculiar velocities of the stars, i.e.~the
quadratical sum of the dispersions of the three velocity components, are
plotted versus the ages of the stars. For this purpose we have extracted three
data sets from the CNS4. The first set includes all main sequence stars down to
late type K stars with known space velocities. These have been grouped
according to $B-V$ colour and the age of each group is assumed to be one half
of
the average main sequence life time of the stars, or for long lived late type
stars one half of the age of the galactic disk, for which we have adopted a
value of
10 Gyrs (cf.~also Jahrei{\ss} et al.~1998, 1999). The second set of stars is
given by the  McCormick K and M dwarfs with known space velocities. The
McCormick stars have been detected by an objective prism survey (Vyssotsky
1963) and are thus free of kinematical bias. Their ages can be determined
using the emission intensities of the emission kernels of the Calcium H and K
absorption lines, because the emission  intensity decreases with increasing
age of a star. Emission intensities have been indexed by Wilson \&
Woolley (1970) for
almost 300 McCormick stars on a relative scale. Grouping the stars according
to their emission indices and assuming a constant star formation rate allows
one then to estimate the mean age of each group (Wielen 1974). In this way
older age groups with ages larger then 5 Gyrs can be resolved. The third set
of stars consists of 206 main sequence stars with $0.5 \leq B-V \leq 1$
(spectral type
G and K), for which the chromospheric emission has been measured with the
Mt.Wilson H and K spectrophotometer (Soderblom et al.~1991). S--measures can
be transformed according to Noyes et al.~(1984) to $\log{R'_{\rm hk}}$. These
can be used again
as age estimators and have been calibrated quantitatively by Donahue (1998)
on an absolute age scale. Finally the velocity dispersion of young Cepheids is
shown as normalization. All
velocity dispersions shown in Fig.~1 have been calculated by weighting the
velocity components of each star by the absolute value of its vertical
velocity $|$W$|$ and are thus representative for a cylinder perpendicular to
the galactic plane at the Sun's position (Wielen 1974 and cf.~section 3 below).
As can be seen  from Fig.~1 all three independent data sets give a consistent
picture of the rise of the velocity dispersion. The diagram in Fig.~1 can be
read from the right to the left and vice versa. In the former case one relates
the velocity dispersions of the stars to their epoch of formation. They would
reflect then the conditions of the interstellar matter at their birth.
However, one would expect then a different evolution of the velocity
dispersions from what is observed. In particular, the sudden drop of the
velocity dispersions of the youngest stars with ages less than 2 Gyrs would
single out the present epoch, which violates the cosmological principle
(Wielen 1977). Since the work of Spitzer \& Schwarzschild (1951, 1953) it is
generally believed that the diagram has to be read from the left to the
right, i.e.~that the velocity dispersions of the stars depend on the ages of
the stars. The rise of the velocity dispersion reflects then stochastical
accelerations of the stars by massive perturbers, which lead to diffusion of
stars in velocity space. The solid line drawn in Fig.~1 indicates an empirical
fit to the data of the form
\begin{equation}
\sigma_\upsilon = \sqrt{ \sigma_{\upsilon_0}^2 + C \tau},
\end{equation}
with $\tau$ the age of the stars.
\begin{figure} [h]
\plottwo{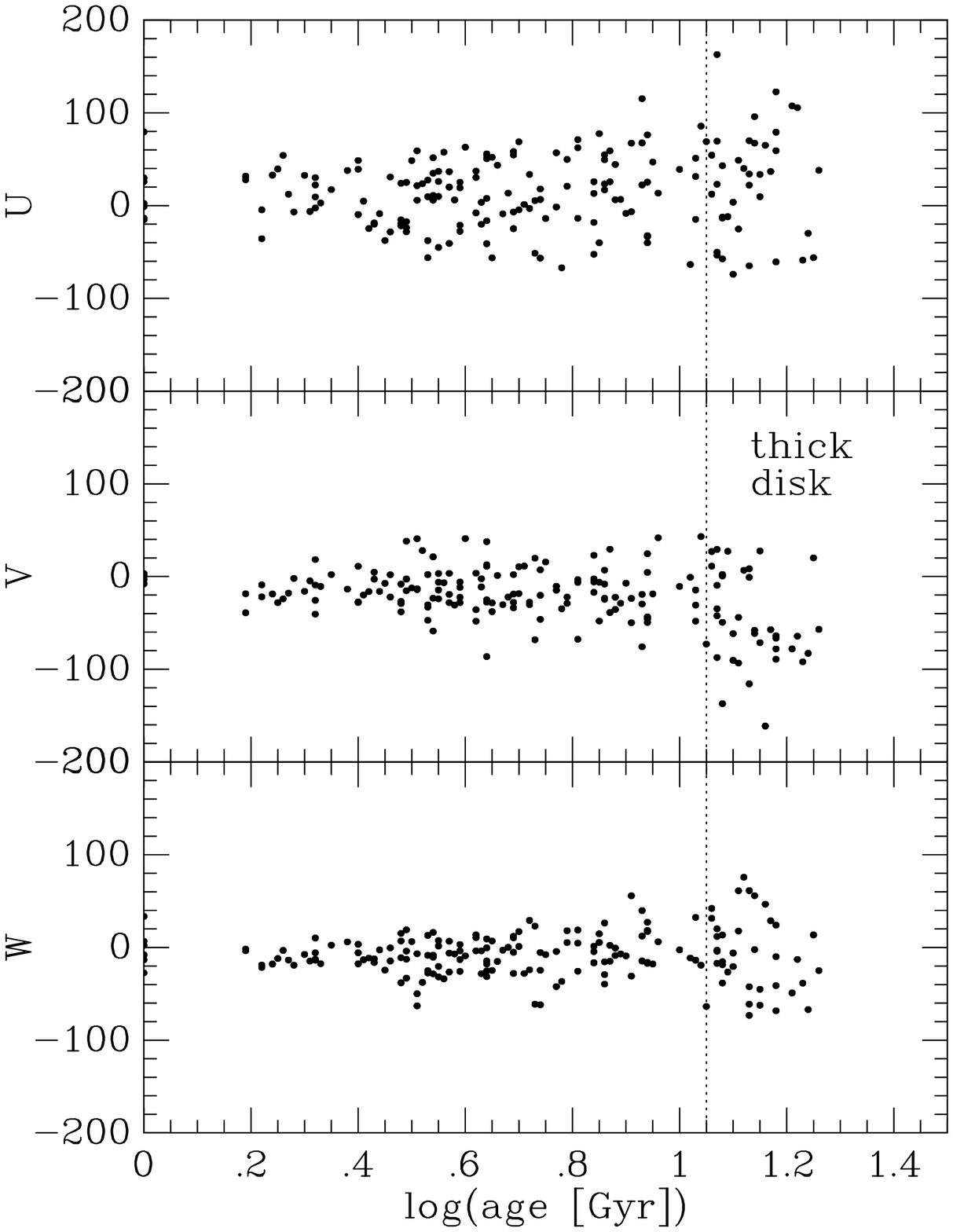}{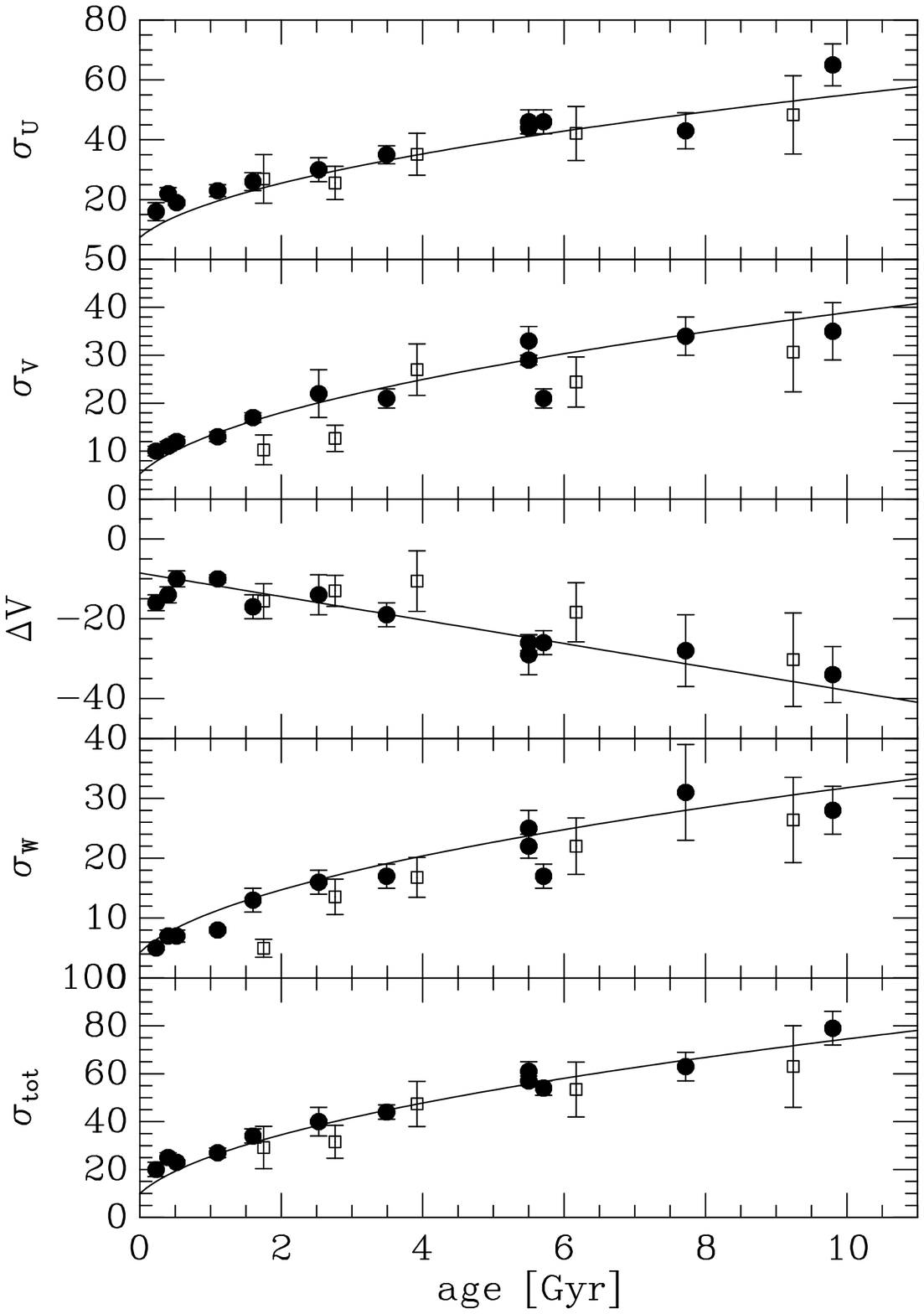}
\caption{Left panel:
Scatter plot of the velocities of the stars from the Edvardsson et al.~(1993)
sample. The U velocity components point to the galactic center, the
V components into the direction of galactic rotation, and the W components
toward the galactic pole, respectively. The vertical dotted line indicates the
transition from old thin disk to thick disk stars.\\ Right panel:
Velocity dispersions of the stars from the Edvardsson et al.
sample (squares) in comparison with the CNS4 data (full dots).
From top to bottom the radial and tangential velocity dispersions, the
mean rotational velocity (with respect to the Sun), the vertical and total
velocity dispersions are shown, respectively. Fits to the velocity dispersion
data
according to equation (1) are drawn as solid lines, scaling as $\sigma_{\rm
U} : \sigma_{\rm V} : \sigma_{\rm W} : \sigma_{\rm tot} =
3.5 : 1.5 : 1 : 5.5$. The asymptotic drift relation is given by  $\Delta {\rm
V} = -8-0.01 \sigma_{\rm U}^2$.}  \label{fig2}
   \end{figure}

\section{
Comparision of CNS4 data with other modern data sets}

We have compared our kinematical data with the data of the Edvardsson et al.
(1993) sample of nearly 200 F and G stars with measured space velocities and
very accurately determined metallicities. In Fig.~2 the Edvardsson et al.~data
are shown as scatter plots. The gradual widening of the distributions with
increasing age of the stars can be clearly seen. Beyond 11 Gyrs the
distributions widen abruptly and the mean rotational velocity of the stars
lags about 40 km/s behind that of the younger stars. This is typical for stars
of the thick disk and we interpret this abrupt change as the transition from
the old thin disk to the thick disk. The separation of thick disk
from old thin disk stars found here is consistent with other studies of
the thick disk (cf.~Gilmore's article in this volume), which also show that
the thick disk is old. In the following we concentrate on the dynamical
evolution of the old thin disk.

We have grouped the Edvardsson et al.~(1993) stars into age groups and have
calculated $|$W$|$--weighted velocity dispersions for each group. These are
also shown in Fig.~2 overlaid over the CNS4 data. For this purpose we have
increased the age scale of the solar neighbourhood data by 10\% to that of
Edvardsson et al.~(1993). From the comparison we conclude that both data
sets are entirely consistent with each other.

\begin{figure}
\plottwo{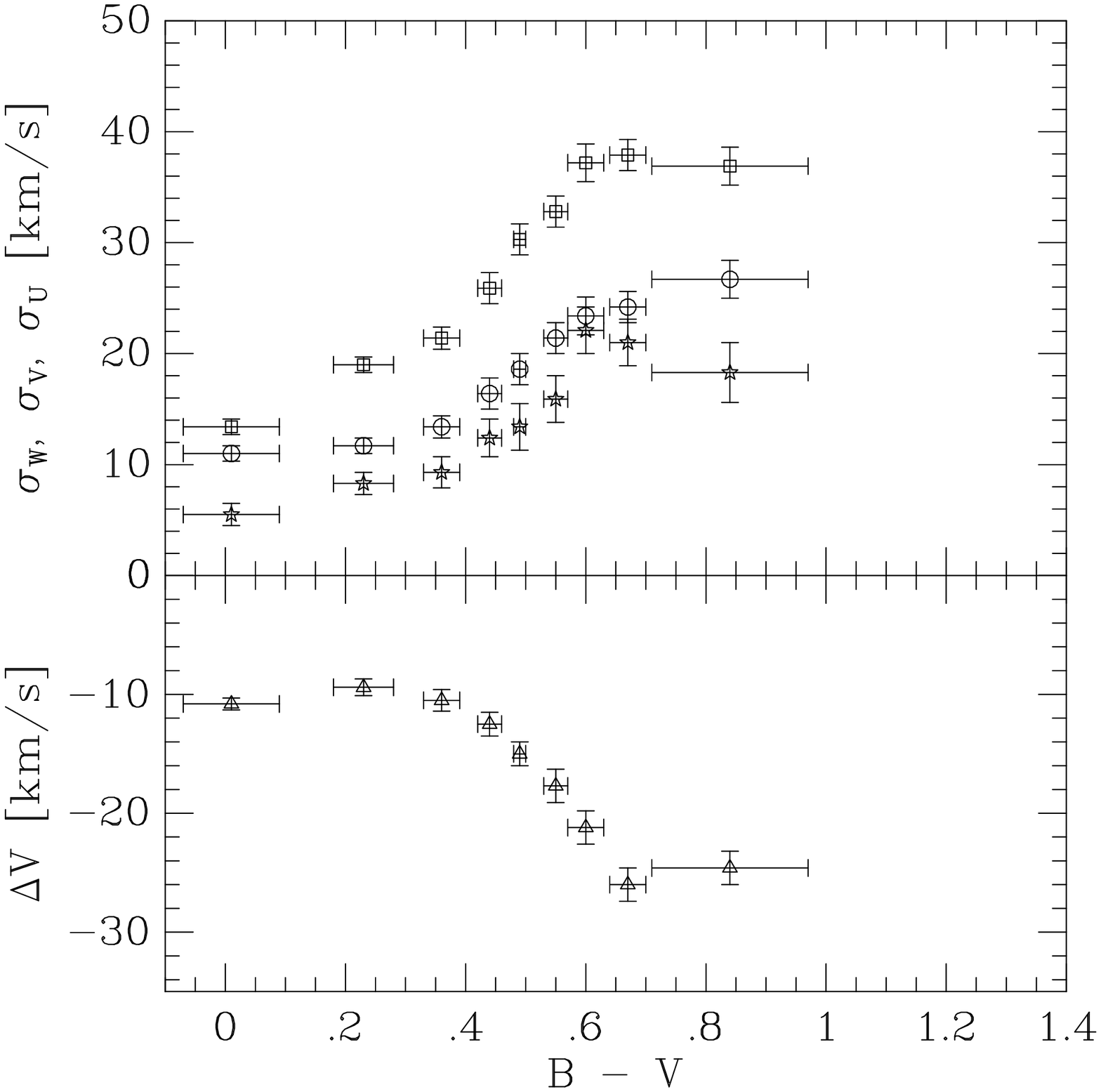}{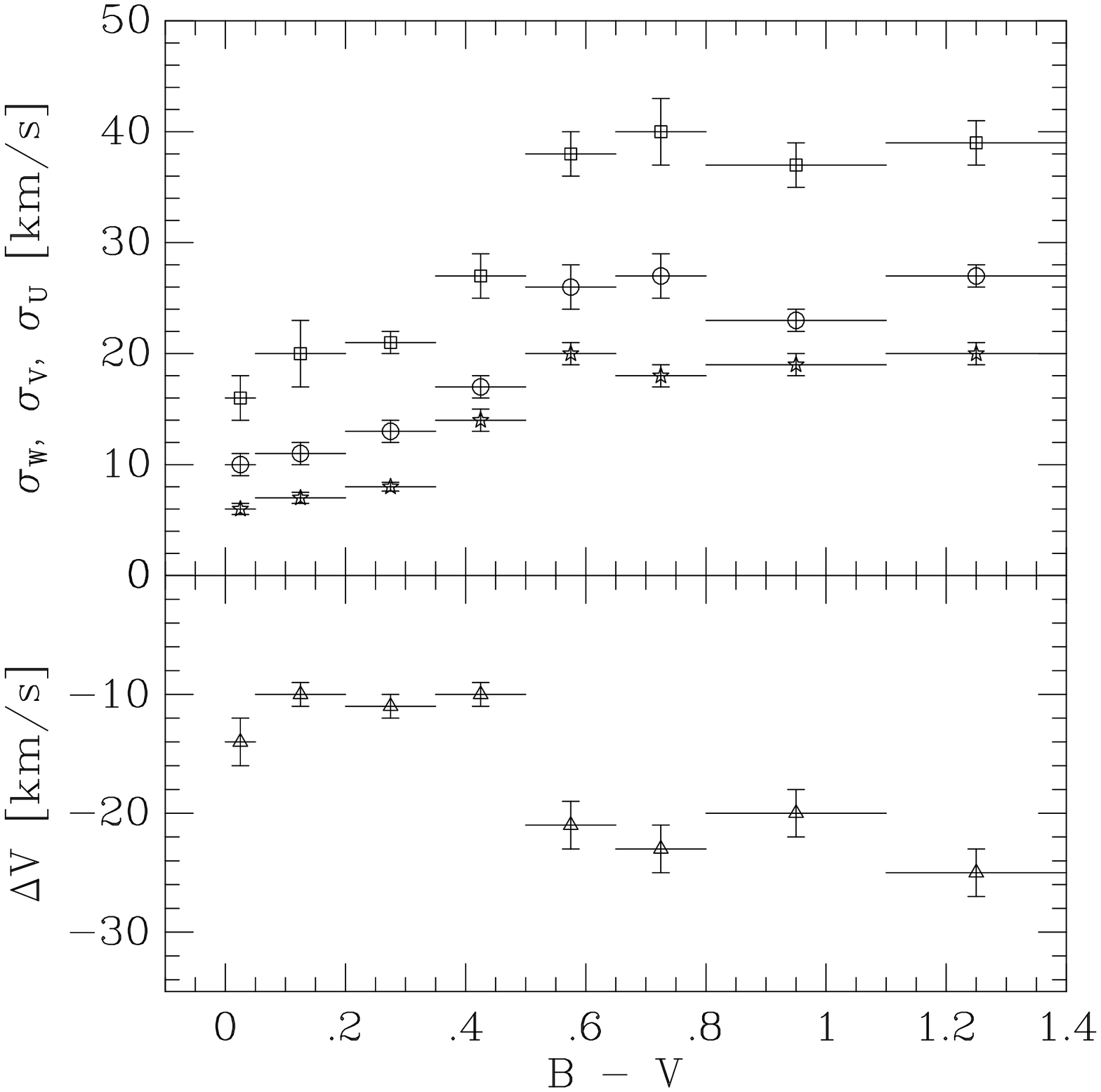}
\caption{Left panel:
Velocity dispersions and rotational lag (with respect to the Sun)
of the stars from the Dehnen \& Binney (1998) sample as function of the $B-V$
colours of the stars. Radial velocity dispersions are indicated by squares,
tangential velocity dispersions by circles, and vertical velocity dispersions
by asterisks, respectively.\\ Right panel: CNS4 data}
 \label{fig3}
   \end{figure}

Recently Dehnen \& Binney (1998) have constructed a new important sample of
kinematical data. By careful comparison of the Hipparcos and Tycho
catalogues (ESA 1997) they were able to extract from the Hipparcos catalogue
a kinematically unbiased sample of more than 14\,000 stars for which reliable
parallaxes and proper motions are available. Dehnen \& Binney (1998) had no
radial velocities available, but making use of the many thousand viewing angles
through the velocity ellipsoid they determined velocity dispersions of groups
of stars grouped according to their $B-V$ colour. These are reproduced in
Fig.~3, where also the corresponding CNS4 data are shown, which refer now to the
local volume and have been thus not weighted by $|$W$|$. The interpretation of
the diagrams is that stars bluer than $B-V$ = 0.6 are younger than the age of
the disk, so that the rise of the velocity dispersions and the increasing
rotational lag reflect the dependance of the velocity dispersions and the
rotational lag on the ages of the stars. Beyond $B-V$ = 0.6 the main sequence
life times of the stars are larger than the age of the disk and the curves
level off at Parenago's discontinuity. It should be kept in mind that a group
of stars of given colour always contains a mixture of young and old stars.
As Fig.~3 shows also the Dehnen \& Binney (1998) data
are in perfect agreement with the solar neighbourhood data. More recently
Binney et al.~(2000) have derived from their data a velocity dispersion --
age relation making use of the Padua isochrones and experimenting with various
star formation histories.
Their preferred result is shown in Fig.~4 in comparison with the solar
neighbourhood data for which again the age scale has been increased by 10\%.
Formally the Binney et al.~(2000) relation is described by a $\sigma_{\rm tot}
\propto \tau^{0.33}$ law, while the solar neighbourhood data  are better
described by a $\sigma_{\rm tot} \propto \tau^{0.5}$ relation. But the
implied discrepancy is only very mild. In particular Binney et al.~(2000)
confirm the high values of the velocity dispersion of the old disk stars.
Rocha--Pinto (priv.~comm.) has analyzed yet another sample of late type stars 
and finds a velocity dispersion -- age relation, which is also similar to the 
CNS4 data.

\begin{figure} [h]
\begin{center}
\epsfxsize=9cm
   \leavevmode
     \epsffile{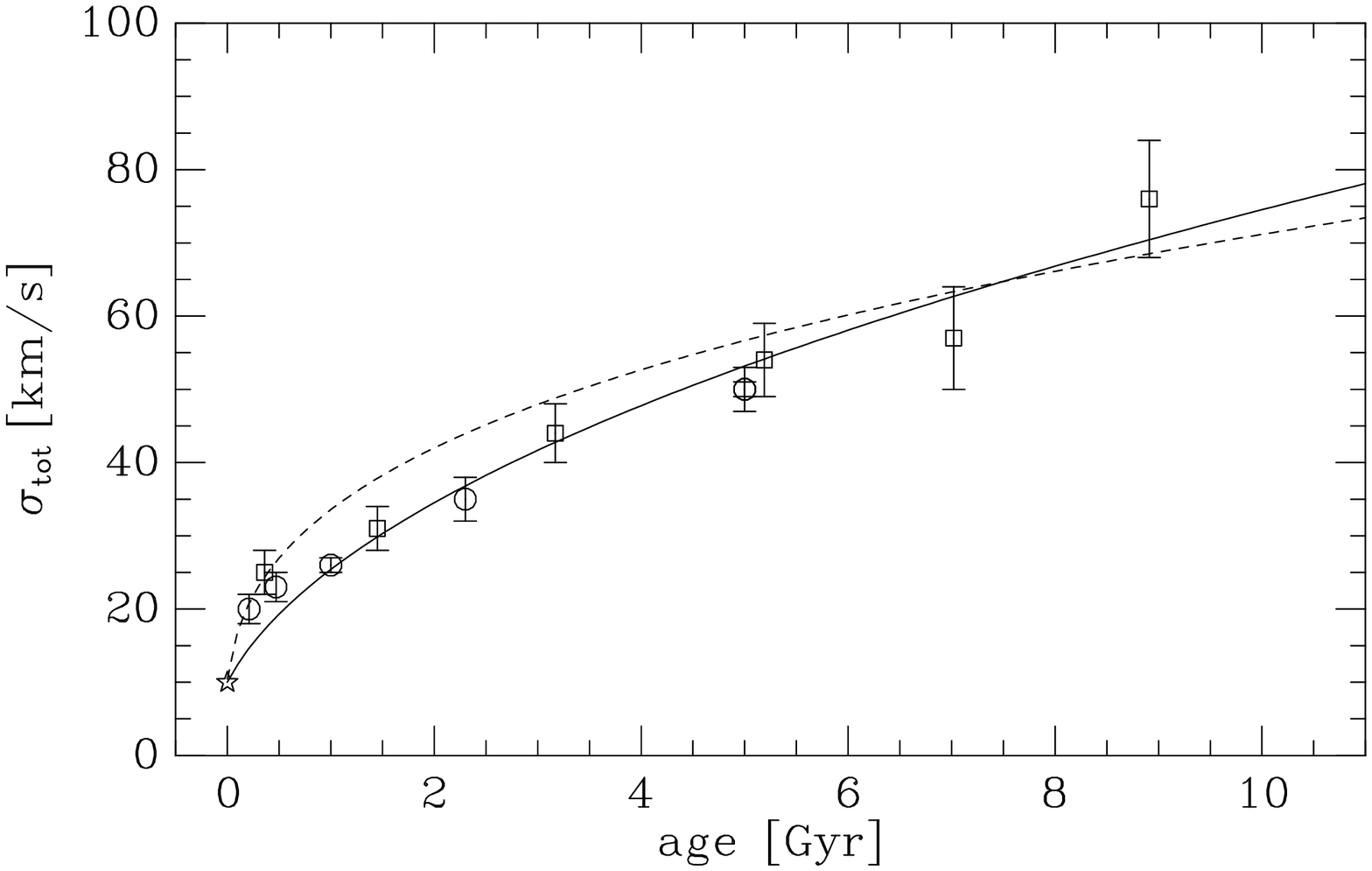}
\caption{
Total velocity dispersion as function of stellar age according to
Binney et al.~(2000) (dashed line) in comparison with the solar
neighbourhood data (solid line and symbols coded as in Fig.~1).}
 \label{fig4}
   \end{center}
   \end{figure}
   
Taking all this together, we conclude that all modern data sets on the
kinematics of nearby stars are in excellent agreement. Finally we note that
very recently Quillen \& Garnett (2000) have presented a velocity dispersion
-- age relation, which is at variance with the results described here. This
appears to be related mainly to the V velocity dispersions derived by them and
is at present not understood.

\section{Star formation history traced in the solar neighbourhood}

The sample of G and K dwarfs from the CNS4 with individually determined ages
using their chromospheric HK emission fluxes as described in section (1) can
be used to trace the star formation history of the galactic disk. The star
formation rate is defined per vertical column density. Thus to each star
detected
in the local volume a weight has to be assigned, which accounts for the number
of similar
stars expected in the cylinder perpendicular to the galactic plane
at the Sun's position. Obviously the weight must be inversely proportional to
the detection probability, which is given by the crossing time of the
star through
the local volume divided by the period of the vertical oscillation of the
star. If the vertical oscillations are assumed to be harmonic, the oscillation
periods of all stars are the same, and the weight is simply given by the
absolute value of the vertical midplane velocity $|$W$|$.
In Fig.~5 a cumulative diagram of the weights
is shown, which illustrates the growth of the surface density of the disk with
time. The slope of the curve is the star formation rate. As can be seen from
Fig.~5 there are hardly any stars older than 7 Gyrs, for which apparently the
chromospheric flux dating cannot be applied. For the younger stars the star
formation rate is fairly constant with a break at an age of the stars of 3.5
Gyrs. The ratio of the slopes is 1 : 3, indicating an enhanced star formation
rate at earlier epochs of the evolution of the galactic disk. Due to the
non-linear K$_{\rm z}$ force law the
actual vertical oscillations of the stars are not harmonic. We have performed
numerical integrations of stellar orbits in a realistic model of the potential
of the galactic disk as determined by Flynn \& Fuchs (1994). The resulting
oscillation periods depend on the vertical midplane velocity as
\begin{equation}
T(|W|) = T_0 \cdot (1 + 0.016 |W|).
\end{equation}
If this correction to the weights is taken into
account, the break in the star formation becomes more pronounced with a ratio
of slopes of 1 : 3.7\,.

\begin{figure} [h]
\begin{center}
\epsfxsize=9cm
   \leavevmode
     \epsffile{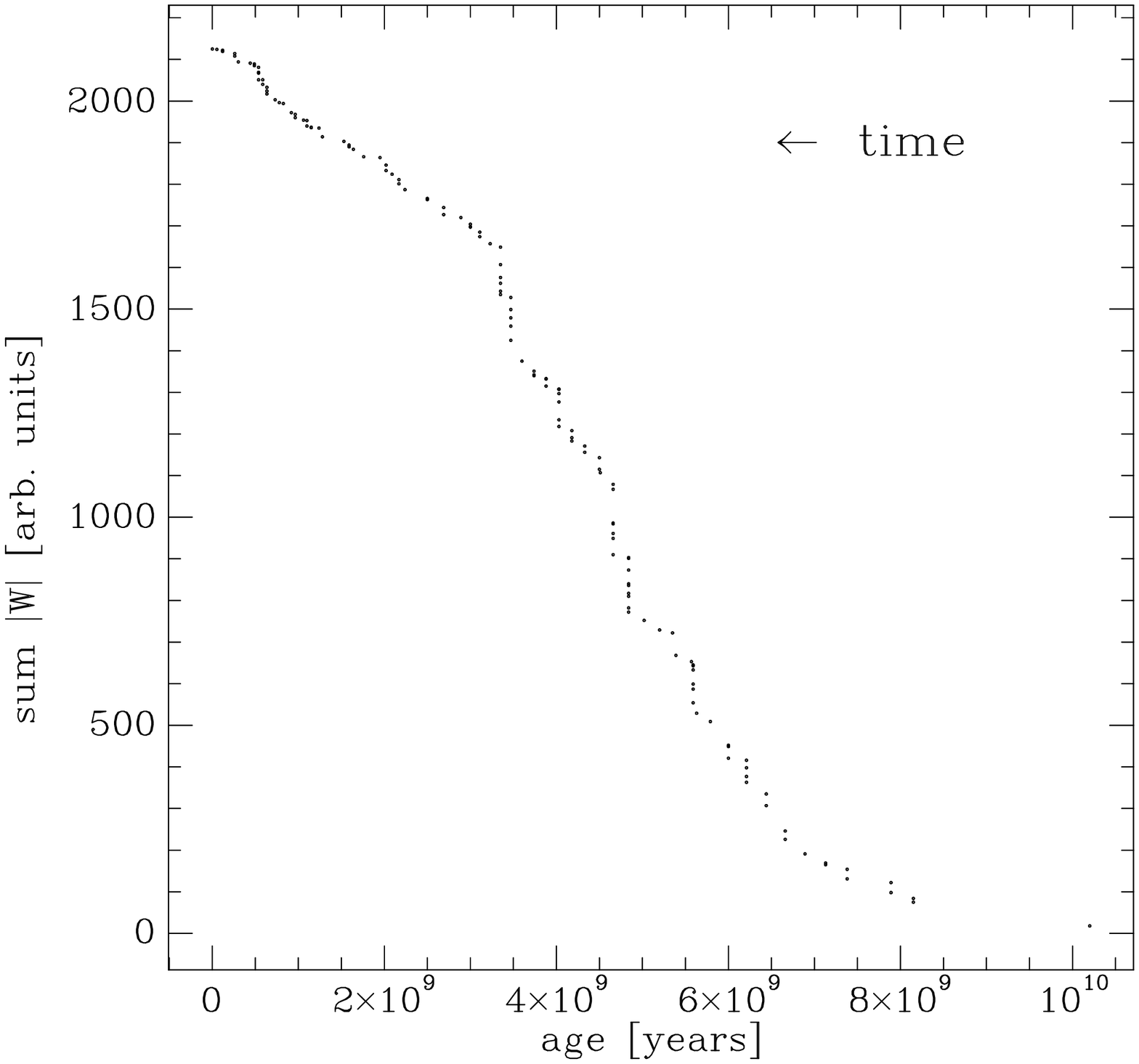}
\caption{
Cumulative star formation rate traced in the solar neighbourhood by a
sample of G and K stars with individually determined ages based on their
chromospheric emission fluxes. The rise of the curve from the right to
the left indicates the growth of the surface density of the galactic
disk during its evolution (see text for details).}
 \label{fig5}
   \end{center}
   \end{figure}

Star bursts appear as `steps' in the cumulative diagram. Hernandez et
al.~(2000)
have applied their method to reconstruct the star formation history to a
sample of bright stars in the solar neighbourhood. They were able to trace the
star formation history back over 3 Gyrs and found three bursts of star
formation at $\tau$ = 0.5, 1.3, and 2 Gyrs, respectively.
These can be seen also distinctly in the cumulative diagram in Fig.~5.
There are also `steps' at earlier times, which are probably related to
star bursts in the evolution of the galactic disk. They correlate, however,
only in part with the star burst periods proposed by Rocha--Pinto et al.~(2000).

\section{Disk heating mechanisms}

Since the pioneering papers by Spitzer and Schwarzschild (1951, 1953) there
is a very rich literature on disk heating, which can not be reviewed here
in its entirety. The many proposed disk heating mechanisms can be grouped
loosely in the following way:

\noindent 1.
Fast massive perturbers of the disk from the galactic halo\\ Such objects
deflect disk stars by gravitational encounters, when they pass through the
disk, and thus heat the disk stochastically. The diffusion coefficients,
which describe the resulting diffusion of the stars in velocity space
quantitatively, are proportional to the typical length of an encounter.
This is inversely proportional to the typical relative speed between disk
stars and massive perurbers. For massive perturbers from the halo this relative
speed is dominated by their velocities and one expects a diffusion
coefficient, which is independent of the speed of the stars (Wielen 1977),
\begin{equation}
\frac{d \sigma^2}{d \tau} = const.,
\qquad {\rm implying} \qquad
\sigma^2 \propto  \tau\,.
\end{equation}
Such type of disk heating mechanisms lead to the velocity dispersion--age
relation, which we have shown in section 1 fitted to the data.
Massive black holes have been proposed as candidates for such kind of
perturbers (Lacey \& Ostriker 1984). This scenario has been given up in the
meantime, however, because such massive black holes, which are thought to
be of primordial nature, would have destroyed the much more fragile disks
of dwarf spiral galaxies (Rix \& Lake 1993, Fuchs et al.~1996).
Recently Moore at al.~(1999) have shown by extensive numerical simulations
that, if the dark halos were assembled according to the cold dark matter (CDM)
cosmology, they would be
highly clumped. According to their simulations about 10\% of the dark matter
is clumped to lumps with typical masses of 5$\cdot$10$^8$ M$_\odot$ and sizes
of 10 kpc
(Moore, priv.~comm.~). The core radii of the lumps are somewhat uncertain,
because they are of the order of the spatial resolution of the simulations,
typically 1 kpc. We have run numerical simulations of the disk heating effect
due to such lumps using a sliding grid scheme with periodic boundary
conditions (Fuchs et al. 1994). The central grid frame represents a patch
of the galactic disk and is populated by test stars, which are refed into
the central frame, when they leave this, as if entering from neighbouring
grid frames, which slide along the central frame (see also the contribution by
H\"anninen \& Flynn in this volume). This model disk is then exposed to
perturbations by the dark matter lumps modelled as Plummer spheres on
isotropic orbits with parameters
as found by Moore et al.~(1999). The resulting disk heating effect is
illustrated in Fig.~6. By varying the core radii we found that the amplitude
of the growth of the velocity dispersion depends very much on the adopted
core radii and is, thus, at present not well constrained. Striking,
however, are the large bulk motions of the stars induced by the
dark matter perturbations. By analyzing the simulations we could
identify each of the sudden jumps of the mean velocities of the
test stars with the passage of a perturber through the
model disk. The implied mean velocities are much larger than
actually observed, and would present a severe challenge to
CDM cosmology, if the results of Moore et al.~(1999) are indeed correct.
\begin{figure}
     \plotone{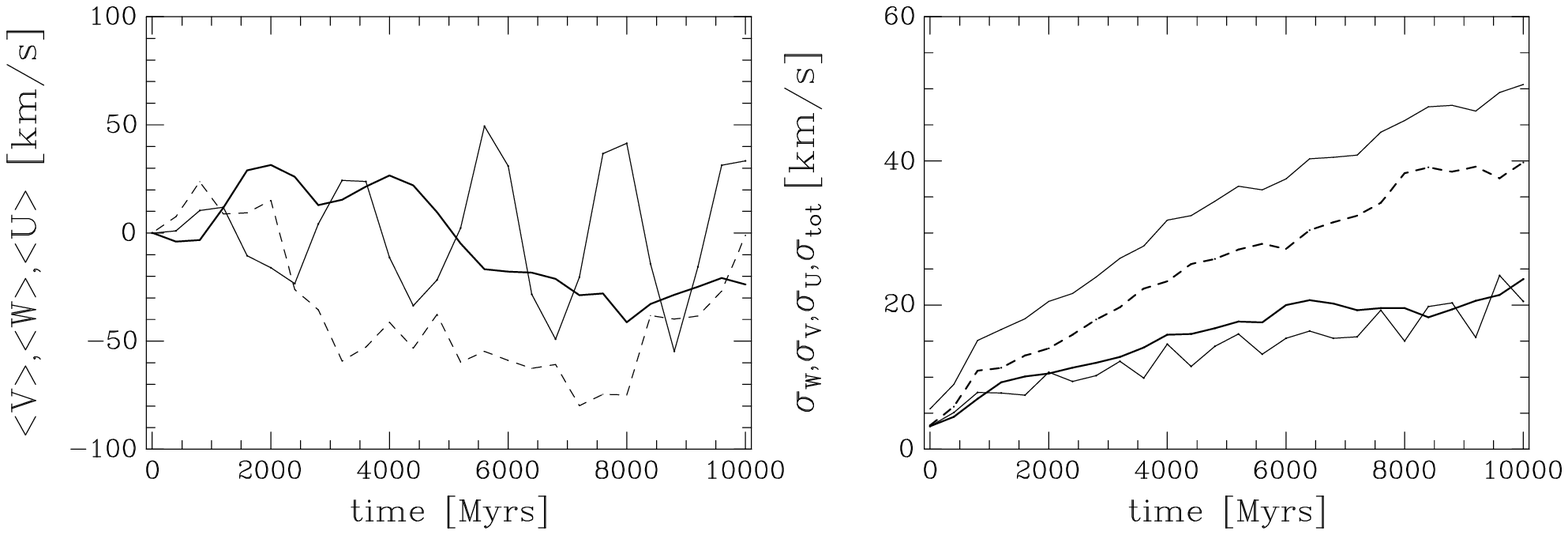}
\caption{
Results of the numerical simulation of the effects of the bombardement
of the galactic disk with massive cold dark matter clumps from the galactic
halo. The left panel shows the resulting mean velocities ($<$U$>$, $<$W$>$,
$<$V$>$ from top to bottom over the first 1000 Myrs) and the right panel
shows the resulting velocity dispersions ($\sigma_{\rm tot}, \sigma_{\rm
U}, \sigma_{\rm V}, \sigma_{\rm W}$ from top to bottom).}
 \label{fig6}
   \end{figure}

\noindent 2.
Slow massive perturbers of the disk such as molecular clouds\\ In that
case the diffusion coefficients are inversely proportional to the typical
stellar peculiar velocity, because the molecular clouds are essentially at
rest in the corotating frame,
\begin{equation}
\frac{d \sigma^2}{d \tau} = \frac{const.\,}{\sigma}, \qquad {\rm implying}
\qquad \sigma^2 \propto  \tau^{1/3}\, .
\end{equation}
Actually the molecular clouds form a thin layer, and since the disk stars spend
most of the time out of this layer, the disk heating becomes even more
ineffective (Lacey 1984), $\sigma^2 \propto  \tau^{1/4}$.
In one of the panels of Fig.~7 we show the result of a numerical simulation
with the sliding grid scheme described above of the disk heating effect due
to molecular clouds. The adopted parameters are: mass of a molecular cloud
5$\cdot$10$^5$\,M$_\odot$, Plummer radius 20\,pc, the vertical scale height
of the cloud layer 55\,pc, and surface density of the molecular clouds
2.5\,M$_\odot$\,pc$^{-2}$. As can
be seen from Fig.~7 the disk heating effect due to molecular clouds is
in the solar neighbourhood negligible. It should be kept in mind,
however, that molecular clouds are very effective in deflecting
stellar orbits (Lacey 1984). If the main heating
mechanism affects mainly the planar velocity components like density waves,
for example, the molecular clouds establish the W velocity distribution.

\noindent Finally we only enumerate:

\noindent 3.
Transient spiral arms (Sellwood \& Carlberg 1984, Binney \& Lacey 1988)

\noindent 4.
Infalling, disrupting satellite galaxies (see Velazquez and White (1999)
for a recent paper)

\section{Tracing the dynamical stability of the galactic disk}

The criterion of stability of galactic disks against gravitational
instabilities is described  quantitatively by the Toomre stability parameter
\begin{equation}
Q = \frac{\kappa \sigma_{\rm U}}{\alpha G \Sigma_{\rm d}}\,,
\end{equation}
where $\kappa$ denotes the epicyclic frequency of the stellar orbits,
$\sigma_{\rm U}$ the radial velocity dispersion of the stars,
$G$ the constant of gravity, and $\Sigma_{\rm d}$ the surface density of
the disk, respectively. $\alpha$ is a numerical coefficient ranging
from 3.4 for a stellar disk with a Schwarzschild velocity distribution
function (Toomre 1964) to 3.9 for a stellar disk with an exponential
velocity distribution function (Fuchs \& von Linden 1998). For an
isothermal gas disk the coefficient is given by $\alpha = \pi$. As is well
known, the requirement for dynamical stability is
\begin{equation}
Q \geq 1\,.
\end{equation}
Galactic disks are compound systems of the stellar and the interstellar gas
disks. Since the compound disk is more unstable than each disk taken alone,
the requirement for dynamical stability of a compound disk is
\begin{equation}
Q_* > 1, Q_{\rm g} > 1\,.
\end{equation}
Quantitative relations are given, for instance, by Fuchs \& von Linden (1998).
The violation of this condition has severe implications for the dynamics of
the disk. Fuchs \& von Linden (1998) have carried out numerical simulations of
the dynamical evolution of such an unstable compound stellar and interstellar
gas disk. Initially only the gas disk was assumed to be unstable ($Q_*>1,
Q_{\rm g}<1$). Both the stellar and the gas disk react violently to this
condition
by developing many strong short-lived, shearing spiral arms. The ensuing
disk heating effect is very large. The stellar disk heats up within a
Gyr so much that it becomes dynamically totally inactive. Only star
formation of stars on low velocity dispersion orbits can prevent this
rapid heating. The numerical simulations of Fuchs \& von Linden (1998)
indicate that, if the star formation rate is of the order $\dot{\Sigma}_{\rm
d}/\Sigma_{\rm d} \approx 0.3 {\rm Gyr}^{-1}$,
the stellar disk stays cool enough to show well developed spiral structure,
although the velocity dispersion of the old stars is continuously rising.

It is instructive to trace the dynamical stability of the galactic disk back in
time. For this purpose we have back extrapolated the run of the surface
densities
of the stars and the gas as well as the stellar velocity dispersion. The build
up of the surface density of the stars can be determined
using the star formation law $\dot{\Sigma}_{\rm d}$ discussed in section 3,
\begin{equation}
\Sigma_* (t) = \Sigma_* (T) - \int_t^T \dot{\Sigma}_* (t') dt'\,,
\end{equation}
where a present surface density of $\Sigma_* (T)$ = 35 M$_\odot$
pc$^{-2}$ (Holmberg \& Flynn 1998) and a present star formation rate of
0.9 M$_\odot$ pc$^{-2}$ Gyr$^{-1}$ is adopted. The age of the disk is
assumed as T = 11 Gyrs. The star formation rate is back extrapolated according
to Fig.~5 for stars younger than 7 Gyrs and it has been adjusted for
the older stars as described below. The surface density of the gas at earlier
epochs is determined by star formation and accretion of fresh gas,
\begin{equation}
\Sigma_{\rm g} (t) = \Sigma_{\rm g} (T) - \int_t^T (\dot{\Sigma}_{\rm g,accr}
(t') - \dot{\Sigma}_* (t')) dt'\,,
\end{equation}
where the accretion rate is described by an exponential law, $
\dot{\Sigma}_{\rm g,accr} \propto \exp{-t'/t_{\rm a}}$. A present
surface density of cold star forming interstellar gas of
8\,M$_\odot$\,pc$^{-2}$ (Dame 1993) is adopted, while the
present day accretion rate and the accretion
time scale are left as further free parameters.
In the best fitting model shown in Fig.~7 we adopt $\dot{\Sigma}_{\rm g,accr}$
(T) = 0.6 M$_\odot$ pc$^{-2}$ Gyr$^{-1}$
and $t_{\rm a} = 16$ Gyrs. The present day star formation and
accretion rates chosen here are consistent with independent determinations
as tabulated in Thon \& Meusinger (1998). The average velocity
dispersion of the stars at a certain epoch can be determined directly
from the observed velocity dispersion age -- relation,
\begin{equation}
\sigma_{\rm U}(t) = \frac{1}{t} \int_0^t \sigma_{\rm U} (\tau) d \tau\,.
\end{equation}
In order to check the plausibility of the model and to constrain the free
parameters, we have also calculated the evolution of the metallicity $Z$ by
integrating the enrichment equation,
\begin{equation}
\frac{\dot{Z}(t)}{Z(T)} = \frac{y}{Z(T)} \frac{\dot{\Sigma}_*(t)}{\Sigma_{\rm
g}(t)} - \frac{Z(t)}{Z(T)} \frac{\dot{\Sigma}_{\rm g,accr}(t)}{\Sigma_{\rm
g}(t)}\,,
\end{equation}
backwards in time. Equation (11) is based on the instantaneous recycling
approximation and we interpret the metallicity calculated in this way
as usual as representative
for the [O/H] abundance. [O/H] is converted to [Fe/H] by the empirical relation
[Fe/H]=1.43 [O/H] (Sommer-Larsen \& Yoshii 1989). A mean present day
metallicity of 0.05 dex is adopted and a yield of $y/Z_\odot$ = 0.66
(reduced by 10\% to $y/Z(T)$) as suggested by Pagel
(1997) is assumed. From the metallicities and the surface density of the stars
we construct the relative distribution of stars over their metallicity,
which we compare
with the empirical distribution obtained by Wyse \& Gilmore (1995) using data
of G dwarfs from the old thin disk. The G dwarf distribution constrains the
free parameters fairly tightly and the best fitting model is shown in Fig.~7.
As final step we have calculated the Toomre stability parameters for the
stellar and gas disks, respectively. 
\begin{figure}
     \plotone{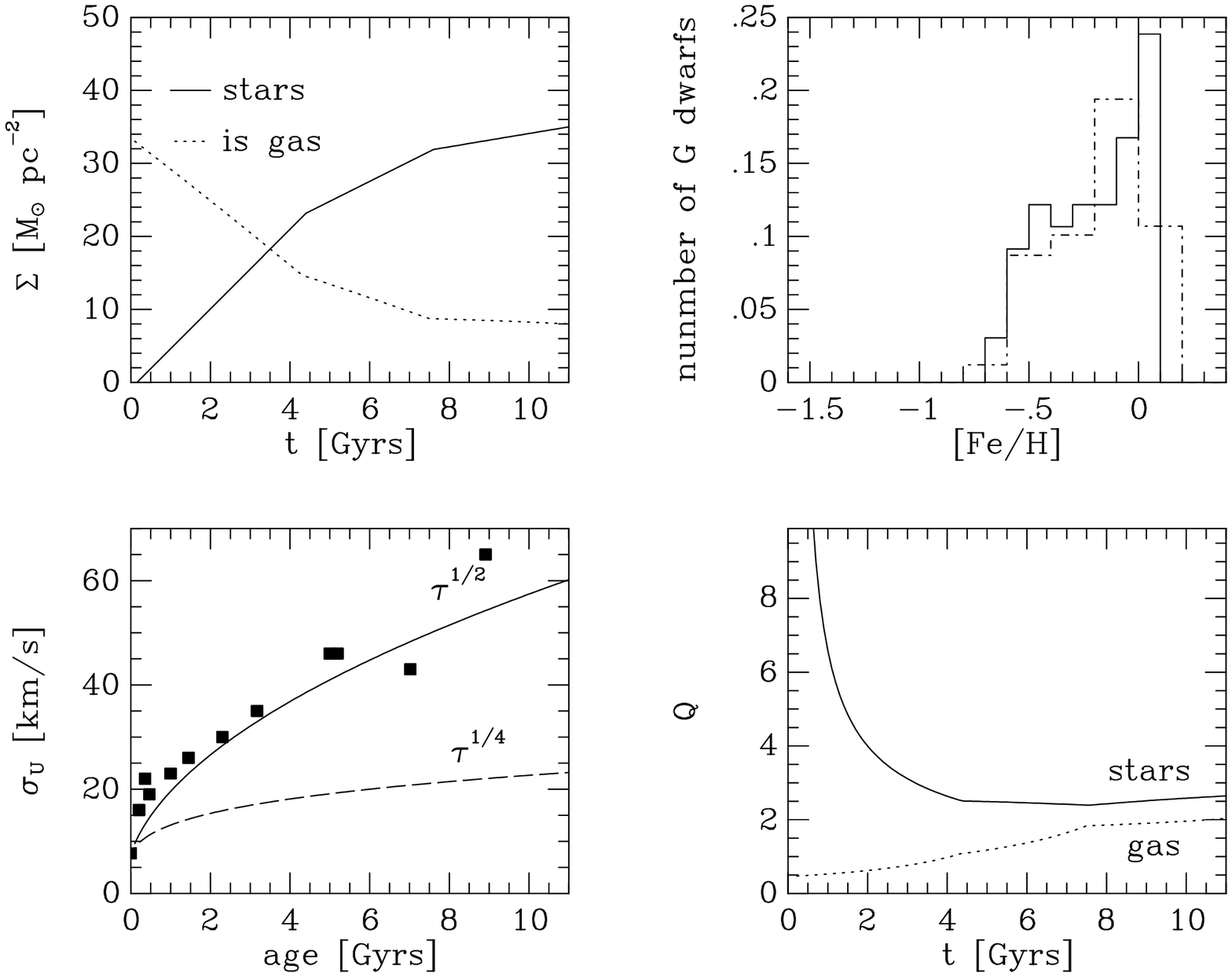}
\caption{
Upper left panel: Surface densities of stars and interstellar gas in
the solar neighbourhood extrapolated backwards. Upper right panel: Relative
distribution of long--lived stars over their metallicity. The solid line is
the model
and the dash-dotted line indicates the empirical distribution. Lower left panel:
Observed velocity despersions fitted by a $\tau^{1/2}$ law (solid line) and
the disk heating effect due to molecular clouds (dashed line). Lower right
panel: Stability parameters of the stellar and gas disk, respectively,
extrapolated backwards.}
 \label{fig7}
   \end{figure}
Interestingly, the gas disk appears
to have been dynamically unstable in earlier epochs of the evolution of
the Milky Way. This meant dynamical instability for the stellar disk
as well and considerable disk heating. Using the criterion of
Fuchs \& von Linden (1998) one can show that the compound star and gas disks
were dynamically unstable until $t$ = 6 Gyrs ($Q_*=2, Q_{\rm g}=1.4$). Note
that the star formation rate was high enough to keep the stellar disk
dynamically active. But even later
on spiral activity accompanied by star formation will have been quite high
and, taken all together, the transition from instability/near instability
may well explain the break in the star formation rate 3 to 4 Gyrs ago.
However, as can be seen from the lower left panel of Fig.~7, where we
illustrate the disk heating effect due to molecular clouds in comparison
with the observed velocity dispersions, even given this dynamical disk
heating mechanism, which might have heated the old stars, there is still
another disk heating mechanism needed to explain the steep rise
of the velocity dispersion of the young stars. As an aside we note finally 
that the star
formation rate of the model of the evolution of the Milky Way disk described
here can be expressed empirically as a Schmidt law, $\dot{\Sigma}_* \propto 
\Sigma_{\rm g}^{1.55}$. 

\acknowledgements
B.~F.~thanks H.~Meusinger and B.~Moore for helpful discussions.

\subsection*{Discussion}
\subsubsection*{Lynden--Bell:}
You do not need a refocussing of stars within a stream. There are two
diffusions
-- the first is the diffusion that moves around the streams but are large scale
gravity perturbations that leave Liouvilles theorem unchanged so we see the
streams intact. The second diffusion is diffusion by small scale gravity fields
between the different stars in any one stream. Evidently this is a much smaller
diffusion.

\subsubsection*{Fuchs:}
That is an interesting suggestion. I should explain that due to the accurate
Hipparcos parallaxes we also see fine structure in the velocity distribution of
the nearby stars. Besides the well known young moving groups
there is also a hint for the old moving group Wolf 630, which does
not show up among the young stars. We do not see the old Hercules moving group,
because the velocity distribution
of the nearby stars is too sparsely populated in that part of velocity space in
order to find statistically significant crowding.
\end{document}